\documentclass[aps,pre,twocolumn,showpacs,amsmath,amssymb,final,letterpaper]{revtex4-1}
\usepackage{graphicx}
\usepackage{subfigure}
\usepackage[pdftex]{hyperref}
\usepackage{epstopdf}

\begin{document}
\title{Propagation dynamics on networks featuring complex topologies}
\author{Laurent H\'ebert-Dufresne}
\author{Pierre-Andr{\'e} No\"el}
\author{Vincent Marceau}
\author{Antoine Allard}
\author{Louis J. Dub\'e}

\affiliation{D\'epartement de Physique, de G\'enie Physique, et d'Optique, Universit\'e Laval, Qu\'ebec (Qu{\'e}bec), Canada G1V 0A6}
\date{\today}
\begin{abstract}
Analytical description of propagation phenomena on random networks has flourished in recent years, yet more complex systems have mainly been studied through numerical means. In this paper, a mean-field description is used to coherently couple the dynamics of the network elements (nodes, vertices, individuals...) on the one hand and their recurrent topological patterns (subgraphs, groups...) on the other hand. In a SIS model of epidemic spread on social networks with community structure, this approach yields a set of ODEs for the time evolution of the system, as well as analytical solutions for the epidemic threshold and equilibria. The results obtained are in good agreement with numerical simulations and reproduce random networks behavior in the appropriate limits which highlights the influence of topology on the processes. Finally, it is demonstrated that, in the absence of degree correlation, our model predicts higher epidemic thresholds for clustered structures than for equivalent random topologies.
\end{abstract}
\maketitle

\section{Introduction}
Description of propagation phenomena has been one of the most prolific field in complex network theory, mostly because of the range of possible applications: epidemic control, spread of information, virus or pollutant propagation in electronic or biological networks \cite{barrat08}. Most analytical models are based on the random network (RN) paradigm: from the point of view of the propagating agent, random networks are seen as identical for every newly infected individual because of their treelike structure (i.e. no loops). This approach has given rise to different descriptions: some are based on a compartmentalisation of nodes according to their state \cite{anderson91}, others on the generating function formalism \cite{newman01, newman02, allard09, noel09} or hybrid descriptions using mean-field theory \cite{volz08, marder07}; yet all approaches are difficult to generalize to real networks for which the RN paradigm rarely applies.

The importance of topology for propagation dynamics \cite{pautasso08, volz08, may06, keeling05, keeling05b, shirley05, newman02}, and more specifically, the importance of clustering \cite{miller09, gleeson09, newman09, ferrari06, newman03a, newman03b}, is now well established. That is, the dynamics on the network is far from independent on how links are arranged between its elements. Furthermore, most real networks feature a significant amount of substructures that simply cannot be ignored as they define the very identity of the networks. The multi-protein units of molecular biology \cite{ravasz02, spirin03}, the coupling of a given set of stocks \cite{onnela03, heimo07} or groups of highly connected individuals \cite{palla07, newman03b} are all good examples of how precise mechanisms (e.g. the friend of my friend is my friend) give rise to important structures within a seemingly random topology.

The two limits of complex networks, complete randomness and perfect order, can be treated with the previously discussed methods. We will concentrate on those particular complex networks, located somewhere between order and disorder, and show how their topology can be taken into account in dynamical problems. In doing so, the language of social networks and epidemics will be used to take advantage of its eloquence and clarity. It should be clear however that the formalism developped is general to many types of networks and propagation phenomena.

The paper is structured as follows. The particular topology chosen to illustrate our approach, the community structure (CS), is described in Sec. \ref{section:CS}. The analytical model is then developed in Sec. \ref{section:main} where we also obtain analytical solutions for the equilibria and epidemic threshold of the system. Section \ref{section:results} compares our analytical results with numerical Monte-Carlo (MC) simulations and presents discussions of our findings. After presenting our conclusions in Sec. \ref{section:conclu}, an Appendix completes our analysis of propagation phenomena on community structure.

\section{Community structure \label{section:CS}}

\begin{figure}[!t]
\centering
\includegraphics[trim = 0mm 108mm 0mm 0mm, clip, width=0.48\textwidth]{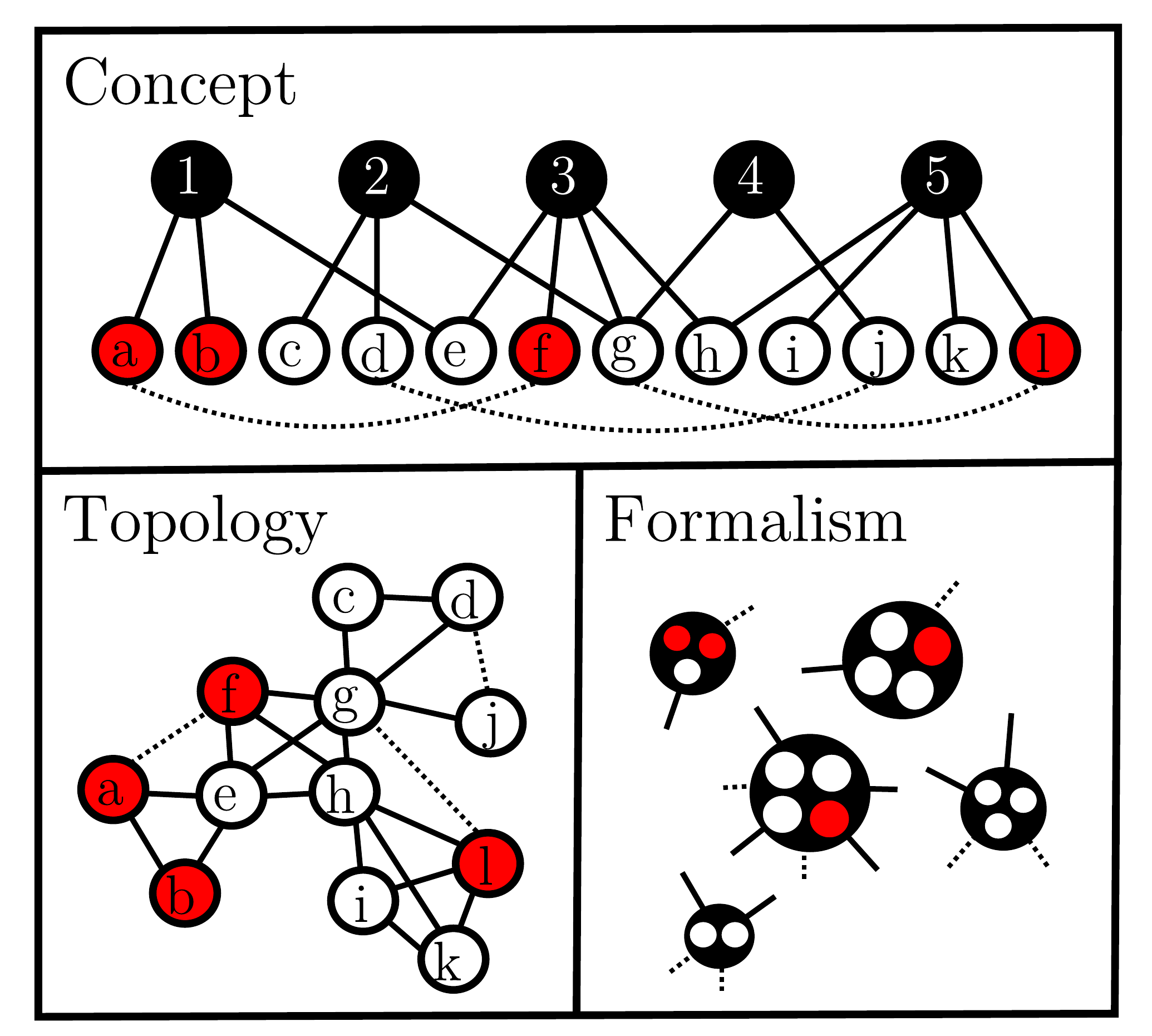}
\caption{(Color online) Schematization of the particular topology studied in this paper. An open mark represents a susceptible individual; a shaded one, an infectious; and a black one, a group (or clique). The topology is constructed by allowing individuals to belong to a given number of cliques where they can be linked to other participants (solid lines) and then randomly assigning random exterior neighbors (dotted lines). Note that in the formalism, the cliques are differentiable by their exact population and state, while the precise connections between them remain unspecified and they are simply linked to a mean-field.}
\label{schema}
\end{figure}

In what follows, a new approach to describe dynamical problems on complex topologies will be used to solve a disease propagation model on social networks featuring a well-known topology: the community structure. We define this particular arrangement of nodes by their aggregation in highly connected groups. These communities (or cliques) can virtually represent a person's family, workplace, collection of friends, etc. This simple concept results in a network with highly connected communities and a sparser density of links between them (see Fig. \ref{schema}). The topology of such networks has been studied at some length: for its initial description, see \cite{girvan02}; for its statistical significance, see \cite{karrer08}; for its detection or characterisation, see \cite{newman06, pujol06, newman04a, newman04b, newman04c}; and the references therein for an exhaustive presentation.

Unfortunately, not unlike other complex types of networks, studies of dynamical processes on this topology has been mainly limited to numerical simulations (e.g. \cite{huang07}). Albeit useful to estimate its effect on the dynamics, they lack the clarity of an analytical framework. On the other hand, mean-field description of propagation phenomena in terms of communities (or households) has been previously attempted in \cite{ball99, ghoshal04, hiebeler06} with several shortcomings such as homogeneous topology, lack of the concept of individuals or inefficient moment closure approximations. Hence, there is a need for an analytical approach that can accurately take into account the many complexities of social networks in order to describe the time evolution of the system. Because community structure typically includes clustering and degree correlation, our formalism will include the coherent contribution of both properties.

A useful model of social topology was published by Newman in \cite{newman03a}. The networks are constructed as follows: each individual belongs to $m$ cliques and each clique holds $n$ individuals, where both $m$ and $n$ are taken from given distributions. Within every clique, each pair of members has a probability $\epsilon$ of being acquainted. Hence, the entire topology is defined by one parameter $\epsilon$ and two probability distributions $\lbrace g_m \rbrace$ and $\lbrace p_n \rbrace$ generated by the following probability generating functions (PGFs):
\begin{eqnarray}
P_0(z) & = & \sum _{n=0} ^\infty p_n z^n \; , \label{P0} \\
G_0(z) & = & \sum _{m=0} ^\infty g_m z^m \; . \label{G0}
\end{eqnarray}
which are simply built from the probabilities $p_n$ and $g_m$ that a random clique or individual will have $n$ participants or $m$ cliques respectively. Similar functions can be defined to generate the probabilities that a random clique of a random individual is shared by $n-1$ other participants or that a random individual in a random clique participates in $m-1$ other cliques. We simply note that these quantities are proportional to $np_n$ or $mg_m$ and thus find our second set of PGFs:
\begin{align}
P_1(z) = \frac{\sum _n np_nz^{n-1}}{\sum _n np_n} = \frac{P_0'(z)}{P_0'(1)} = \nu ^{-1} P_0'(z) \; , \label{P1}  \\
G_1(z) = \frac{\sum _m mg_mz^{m-1}}{\sum _m mg_m} = \frac{G_0'(z)}{G_0'(1)} = \mu ^{-1} G_0'(z) \label{G1} 
\end{align}
where $\nu$ and $\mu$ are respectively the mean numbers of individuals per clique and cliques per individual used to normalize the distributions. Note that the mean of a distributed quantity is simply given by the derivative of the corresponding PGF evaluated at unity. The following topological properties have already been derived in \cite{newman03a} and \cite{newman03b}: degree distribution, size of the giant component, clustering coefficient and degree correlation. Some of these results are used throughout this paper.

Newman's model, although realistic because of its overlapping communities, is strongly limited since links only arise through communities. A node belonging to a single clique does not participate at all in the coupling, while a node belonging to two cliques or more will have a huge influence. Hence, it is hard to describe weakly coupled communities of significant sizes using this particular topology. Consequently, we will introduce a more general description of community structure where exterior random links are also allowed. We simply add a distribution for the number of random links per individual, which is generated by:
\begin{equation}
K_0(z) = \sum _{l=0} ^\infty k_l z^l \; .
\label{K}
\end{equation}
Our networks will thus be defined by the $\epsilon$ probability and three distributions for the numbers of individuals per clique (\ref{P0}), cliques per individual (\ref{G0}) and random links per individual (\ref{K}). Intuition indicates that a large number of networks can be decomposed as basic structures coupled either by sharing nodes, by forced connections or a combination of both. In fact, many of the previously cited papers study networks where nodes belong to a single clique coupled only by random links with the outside world (e.g. \cite{newman04a, gleeson09}). Our general model includes this topology and Newman's original model as special cases.

\section{SIS model of disease propagation on community structure \label{section:main}}
\subsection{Construction of the dynamical model}
The philosophy behind our formalism is to analyze the network simultaneously from two perspectives, i.e. the state of the network is followed from the point of view of recurrent patterns in its topology and of the elements themselves. More precisely, we compartmentalize both the structure and the node ensemble in terms of their relation to one another and couple the two systems to give a complete description of the propagation phenomenon. For social networks featuring community structure, the recurrent patterns are cliques of individuals that can be distinguished by their size and their state. The elements are individuals distinguishable by the number of cliques to which they belong and by their number of exterior random links. That is, the mean state of a given class of individuals will act as if all of their cliques and random links were approximated by a mean-field and the mean state of a given class of cliques will act as if all individuals were also reduced to a mean-field approximation. The behaviors of both cliques and individuals are coupled in terms of their connections via the generating functions \eqref{P0} through \eqref{K}.

The particular case under study is a Susceptible-Infectious-Susceptible (SIS) model of disease propagation. In continuous time, an infectious node may pass the disease to any of its susceptible neighbors at a rate $\tau$ \mbox{(S $\rightarrow$ I)}, while it is recovering from the disease at a rate $r$ \mbox{(I $\rightarrow$ S)}. Given initial conditions, we are interested in developping a system of equations capable of following the state $I(t)$ of the network, where $I(t)$ is the fraction of infectious individuals at a given time. According to our philosophy, we thus need to follow both individuals and cliques. Let $S_{m,l}(t)$ be the proportion of individuals which belong to $m$ cliques, have $l$ random links and are susceptible at time $t$ and $C_{n,i}(t)$ be the proportion of cliques whose population is $n$ and of which $i$ are infectious at time $t$. For the sake of clarity, we will not explicitly mark the time dependence, $(t)$, when it is obvious that the quantity is a dynamical variable.

First, we need to describe how the generating functions $G_1(z)$, $K_0(z)$ and $P_1(z)$ will differ depending on the state of the involved individual. To define the dynamical generating functions, it is possible to either follow the distributions for the susceptibles or the infectious individuals, since $S_{m,l} + I_{m,l} = g_mk_l$. We will follow the susceptibles. We then need the distribution of cliques reached from a susceptible individual of a given clique. This distribution will be affected by $\{S_{m,l}\}$ in the following manner: a random individual has probability $mg_m$ of belonging to $(m-1)$ other cliques, but consequently, only a probability $\sum _l S_{m,l}/g_m$ of being susceptible at time $t$. The reasoning is even simpler for $K_0(z)$ as the distribution is not affected by the knowledge that the individual belongs to at least one clique. We can directly write:
\begin{equation}
\widetilde{G}_1(z;t) = \frac{\sum _{m,l} m S_{m,l} z^{m-1} }{ \sum _{m,l} m S_{m,l} } \; ,
\label{G_1_mod}
\end{equation}
\begin{equation}
\widetilde{K}_0(z;t) = \frac{\sum _{m,l} S_{m,l} z^l}{\sum _{m,l} S_{m,l}} \; ,
\label{K_mod}
\end{equation}
where the tilde denotes that the function generates a distribution which applies to susceptible individuals only. In a similar fashion, the knowledge that a clique is reached by a link emerging of a susceptible individual will affect the distribution of this clique's number of susceptible individuals. The probability that a susceptible individual belongs to a clique of state $\{ n,i \}$ is directly proportional to the number of susceptible members of that particular state. In order to consider only susceptibles individuals, the $P_1(x,y)$ generating function must be modified accordingly to the number of susceptible members belonging to each compartment:
\begin{equation}
\widetilde{P}_1(x,y;t) = \frac{\sum _{n,i} (n-i)C_{n,i}x^{n}y^{i}}{\sum _{n,i} (n-i)C_{n,i}} \; .
\label{P_1_mod}
\end{equation}
Four interesting and important quantities can be derived from these dynamical generating functions. Firstly, the average number of infectious neighbors per clique and per random link for a susceptible individual, $R(t)$ and $T(t)$:
\begin{eqnarray}
R(t) & = & \epsilon \frac{\sum _{n,i} i(n-i)C_{n,i}}{\sum _{n,i} (n-i)C_{n,i}} \; ,\label{R} \\
T(t) & = & 	\frac{\sum _{n,i} \frac{i}{n}\left(nC_{n,i}\right)}{\sum _{n,i} nC_{n,i}} \; , \label{R2}
\end{eqnarray}
Secondly, the mean number of excess infectious neighbors per clique and per random link for a susceptible individual of a given clique, $\rho (t)$ and $\sigma (t)$:
\begin{eqnarray}
\rho (t) & = & \widetilde{G}'_1(1;t)R(t) \; , \label{rho} \\
\sigma (t) & = & \widetilde{K}_0'(1;t)T(t) \label{sigma} 
\end{eqnarray}
where the primes denote a derivative with respect to $z$, so that $\widetilde{G}'_1(1;t)$ is the average number of outside cliques for a susceptible member of a given clique at time $t$.

Let us now construct the differential equation governing $\{ S_{m,l} \}$. We previously mentionned that the disease spreads through any link between a susceptible and an infectious individual. Thus, with $R(t)$ being the average number of such links that a susceptible may have in a single clique, the rate at which the class of individuals belonging to $m$ cliques is infected, is proportionnal to $-\tau m S_{m,l}R(t)$. Similarly, with $T(t)$ being the probability that a random link leads to an infectious individual, the rate of infection for individuals with $l$ random links must be proportionnal to $-\tau l S_{m,l}T(t)$. Simultaneously, the same ratio increases as the infected nodes recover at a speed $r(g_mk_l-S_{m,l})$. Therefore, the set of equations governing the point of view of the individuals is simply obtained by summing the contributions from these three processes:
\begin{equation}
\dfrac{dS_{m,l}}{dt} = r(g_mk_l-S_{m,l})-\tau S_{m,l} \left[mR(t) + lT(t)\right] \;.
\label{S_ml}
\end{equation}
Similar considerations are needed to define the dynamics of the $C_{n,i}$ values. A clique in a $\lbrace n,i \rbrace$ state can either pass to $\lbrace n,i+1 \rbrace$ by infection (if $i < n$) or to $\lbrace n,i-1 \rbrace$ by recovery (if $i > 0$). The first process is proportionnal to the sum of the number of links between infectious and susceptible individuals within the cliques and the number of links with infectious neighbors that each susceptible might have outside the considered clique. For a given $\lbrace n,i \rbrace$ compartment, infection can either bring new cliques from the $\lbrace n,i-1 \rbrace$ state or cause the cliques to pass to the more infectious $\lbrace n,i+1 \rbrace$ compartment:
\begin{align}
\dfrac{dC_{n,i}}{dt} \propto &\phantom{+}\tau\left(n-i+1\right)\left[\epsilon \left(i-1\right)+\rho (t) + \sigma (t)\right]C_{n,i-1} \nonumber \\
&-\tau \left(n-i\right)\left[\epsilon i+\rho (t)+ \sigma (t)\right]C_{n,i} \; .
\label{C_ni_1}
\end{align}
The contribution of the recovery process is easy to explicit using the same logic, as it is simply proportionnal to the number of infectious individuals who might recover:
\begin{align}
\dfrac{dC_{n,i}}{dt} \propto r\left(i+1\right)C_{n,i+1} -riC_{n,i} \; .
\label{C_ni_2}
\end{align}
Summing the contributions of both the infections (\ref{C_ni_1}) and the recoveries (\ref{C_ni_2}) yields the desired differential equation for the cliques dynamics:
\begin{align}
\dfrac{dC_{n,i}}{dt} = &\phantom{+}r\left(i+1\right)C_{n,i+1} -riC_{n,i} \nonumber \\
 &+\tau\left(n-i+1\right) \left[\epsilon \left(i-1\right)+\rho (t)+\sigma (t)\right]C_{n,i-1} \nonumber \\
&-\tau \left(n-i\right)\left[\epsilon i+\rho (t)+\sigma (t)\right]C_{n,i} \; .
\label{C_ni}
\end{align}
where $C_{n,i}$ is defined only for $i \in [0,n]$. Coupled with Eq. (\ref{S_ml}), we now have a complete dynamical system for the state of the network in a SIS model of disease spread. 

If desired, the mean fraction of infectious individuals of a given class of cliques can be obtained in a straightforward manner with:
\begin{equation}
I_n = \sum _i \frac{1}{np_n}iC_{n,i} \; .
\label{I_n}
\end{equation}
It is generally simpler to caracterize the state of the network via the total fraction of infectious, $I(t)$, or susceptible, $S(t)$, individuals. From Eq. (\ref{S_ml}), we directly have:
\begin{align}
S(t) = \sum _{m,l} S_{m,l} \; ; \;\;\;\; I(t) = \sum _{m,l} (1-S_{m,l}) \; .
\label{global_state}
\end{align}
Note that a straightforward evaluation of the global state of the network from $\{C_{n,i}\}$ would be biased because an individual belonging to $m$ cliques would be counted $m$ times more than an individual participating to a single clique.

\subsection{Solution for network stable state}
System (\ref{S_ml}) and (\ref{C_ni}) can be solved as a traditional self-consistent field by looking for a solution in terms of $\rho$ and $\sigma$. Using Eq. (\ref{C_ni}) for the $C_{n,i}$ quantities at equilibrium (i.e. $dC_{n,i}/dt=0$), we obtain the following recursive solution:
\begin{align}
C_{n,i+1}^* = \frac{1}{(i+1)r}\left[\left(f_{n,i}+ri\right)C_{n,i}^* -f_{n,i-1}C_{n,i-1}^* \right]
\label{C_sol}
\end{align}
with $C_{n,i} = 0$ $\forall$ $i \notin [0,n]$, and where we introduce a matrix of infection $\{ f_{n,i} \}$ whose elements depend on the total mean-field $\xi$:
\begin{eqnarray}
f_{n,i} & \equiv & \tau (n-i)(i\epsilon + \xi ^*) \label{fi} \\
\xi ^* & \equiv & \rho ^* + \sigma ^* \; . \label{xi}
\end{eqnarray}
Asterisks will hereafter refer to values at equilibrium. Equation (\ref{C_sol}) can be used to fix the stable values of all the $C_{n,i}^*$ relative to $C_{n,0}^*$, which can then be solved exactly by applying the following topological constraint:
\begin{equation}
\sum _i C_{n,i} = p_n \;\;\; \forall \; t,n \; .
\label{topo_cond}
\end{equation}
Using the equilibrium condition on Eq. (\ref{S_ml}) provides a direct solution for the $S_{m,l}^*$ ensemble:
\begin{equation}
S_{m,l}^* = \frac{rg_mk_l}{\tau \left(mR^* +lT^*\right) + r} \; .
\label{S_sol}
\end{equation}
It is then possible to write $R^*$, $T^*$, $\widetilde{G}_1^*(z)$ and $\widetilde{K}_0^*(z)$ in terms of $\rho ^*$ and $\sigma ^*$ by using (\ref{C_sol}) in (\ref{R}) and (\ref{R2}) while using (\ref{S_sol}) in (\ref{G_1_mod}) and (\ref{K_mod}). A transcendental equation is obtained for $\xi ^*$ by writing (\ref{rho}) and (\ref{sigma}) as:
\begin{align}
\!\!\!\!\!\!\!\!\!\!\!\!\xi ^* = & \left[ \frac{\sum _{m,l} m(m-1) S_{m,l}^*}{\sum _{m,l} mS_{m,l}^*} \right] R^* \nonumber \\
&  \qquad\qquad +\left[ \frac{\sum _{m,l} lS_{m,l}^*}{\sum _{m,l} S_{m,l}^*} \right] T^* \equiv F(\xi ^*) \; ,
\label{trans_eq}
\end{align}
where the dependence on $\xi ^*$ comes from that of $\{ S_{m,l}^* \}$ on $R^*$ and $T^*$ written in terms of $\{ C_{n,i}^* \}$ which are a direct function of $\xi ^*$. Solving for $\xi ^*$ yields a unique non-zero solution fixing $\{ C_{n,i} ^* \}$ which in turn provide the values for $R^*$ and $T^*$. This directly fixes $\{ S_{m,l}^* \}$ using (\ref{S_sol}), and thus the stable state of the network defined by (\ref{global_state}).

Clearly the dynamics is governed by the ratio $\lambda \equiv \tau/r$ and not the individual rates. Therefore, under the transformation to the normalized propagation rate $\lambda$, our model admits a single independent parameter in its dynamics.

\subsection{Solution for epidemic threshold}
\begin{figure}
  \centering
  \includegraphics[width=0.46\textwidth]{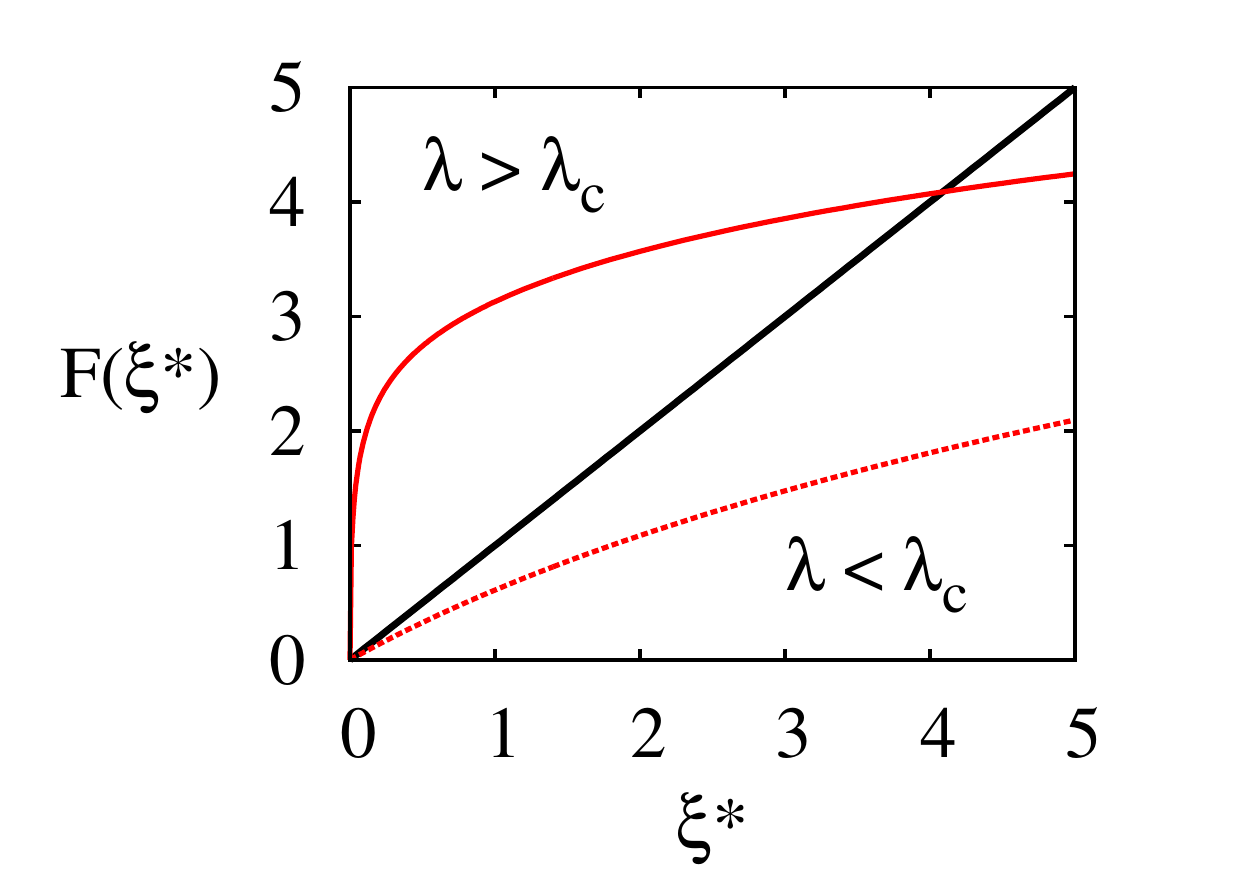}
  \caption{(Color online) Function $F(\xi ^*)$ is shown in shade on the topology defined in (\ref{topo}) for two different normalized propagations rates: $\lambda = 0.02$ in dotted line (under the threshold; no solution for $\xi ^* > 0$) and $\lambda = 0.1$ in solid line (epidemic). The black solid line is the curve of slope 1, $F(\xi ^*) = \xi ^*$.}
  \label{F}
\end{figure}
The epidemic threshold $\lambda_c$ is defined by a phase transition in the normalized infection rate where a macroscopic final epidemic size first appears. Here, it can be defined mathematically using the analytic solution for the stable state of the SIS epidemic. Equation (\ref{trans_eq}) behaves as shown in Fig. \ref{F} with a trivial solution at $\xi ^* = 0$ and another possible solution $\xi ^* > 0$ depending on $\lambda$ and the topology. Since $F(\xi ^*)$ is a monotonously increasing function, $\lambda _c$ can be found by the following condition:
\begin{align}
\frac{d}{d\xi ^*}F(\xi ^*) \bigg\vert _{\xi ^* =0} = 1 \; .
\label{tau_c_cond}
\end{align}
For initial derivative value above unity, a solution $\xi ^* > 0$ exists and the stable epidemic state is non-zero (Fig. \ref{F}). For a system subject to a propagation at its threshold, by definition, we know that the stable state is the trivial solution $C^*_{n,i} = p_n \delta _{i0}$ $\forall$ $\{n,i\}$ and $S^*_{m,l} = g_mk_l$ $\forall$ $\{m,l\}$ (which implies $\widetilde{G}_1(z;t) = G_1(z)$ and $\widetilde{K}_0(z;t) = K_0(z)$). It follows that the mean-field values are zero at equilibrium and (\ref{tau_c_cond}) straightforwardly becomes:
\begin{equation}
\frac{1}{\nu}\sum _{n,i} \bigg\{\epsilon i(n\! -\! i) G'_1(1) + iK'_0(1)\bigg\}\!\frac{d}{d\xi ^*}C_{n,i}^* \bigg\vert _{\xi ^* =0} \!\!\! = 1 \; .
\label{thresh_cond}
\end{equation}
Using (\ref{C_sol}) to evaluate the derivative at equilibrium, one finds that $\forall \; i>0$:
\begin{equation}
\frac{d}{d\xi ^*}C_{n,i}^* \bigg\vert _{\xi ^* =0} = \frac{p_n}{i}\lambda_c^i \epsilon ^{i-1}\frac{n!}{(n-i)!} \; .
\label{thresh_sol}
\end{equation}
Using (\ref{thresh_sol}) to solve (\ref{thresh_cond}) for $\lambda _c$ provides a polynomial with positive coefficients for terms of order one or more:
\begin{equation}
\!\frac{1}{\nu}\!\sum _{n,i>0} \!p_n\!\left(\epsilon \lambda_c\right)^{\! i} \!\frac{n!}{(n\! -\! i )!}\left(\!\!(n\! -\! i)G_1'(1)\! +\!\frac{K_0'(1)}{\epsilon} \right)  = 1 \; .
\label{thresh_poly}
\end{equation} 
This polynomial therefore has a single real positive solution, which is the epidemic threshold of the network. For random networks, one can set $K_0'(1)=0$, $\epsilon = 1$ and $p_n = \delta _{n,2}$ so that all links are shared within cliques of size two. Expression (\ref{thresh_poly}) then reduces to:
\begin{equation}
G_1'(1) \lambda _c ^{\textrm{RN}} = 1 \; ,
\label{tau_c_rg}
\end{equation}
where $G_1'(1)$ is here the mean excess degree. From Eq. (\ref{tau_c_rg}), one can deduce that our model predicts a null SIS epidemic threshold only if $G_1'(1)$ diverges. For scale-free networks whose degree distribution falls as $k^{-s}$, it can be shown that $G_1'(1)$ diverges if $s \leq 3$. Our model therefore leads to the same conclusion as \cite{vespignani01}: scale-free networks with degree distribution $p_k \propto k^{-s}$ and $s \leq 3$ are defined by an absence of epidemic threshold.

A calculation of the SIS epidemic threshold on random networks was previously done in \cite{parshani10}, using discrete time steps and constant recovery period approximations. To the best of our knowledge, Eq. (\ref{thresh_poly}) is the first equation for a continous time SIS model of epidemic spread for both random networks and community structure.

\section{Implementation and validation \label{section:results}}

\begin{figure}[t!]
\centering
\includegraphics[trim = 0mm 0mm 0mm 0mm, clip, width=0.46\textwidth]{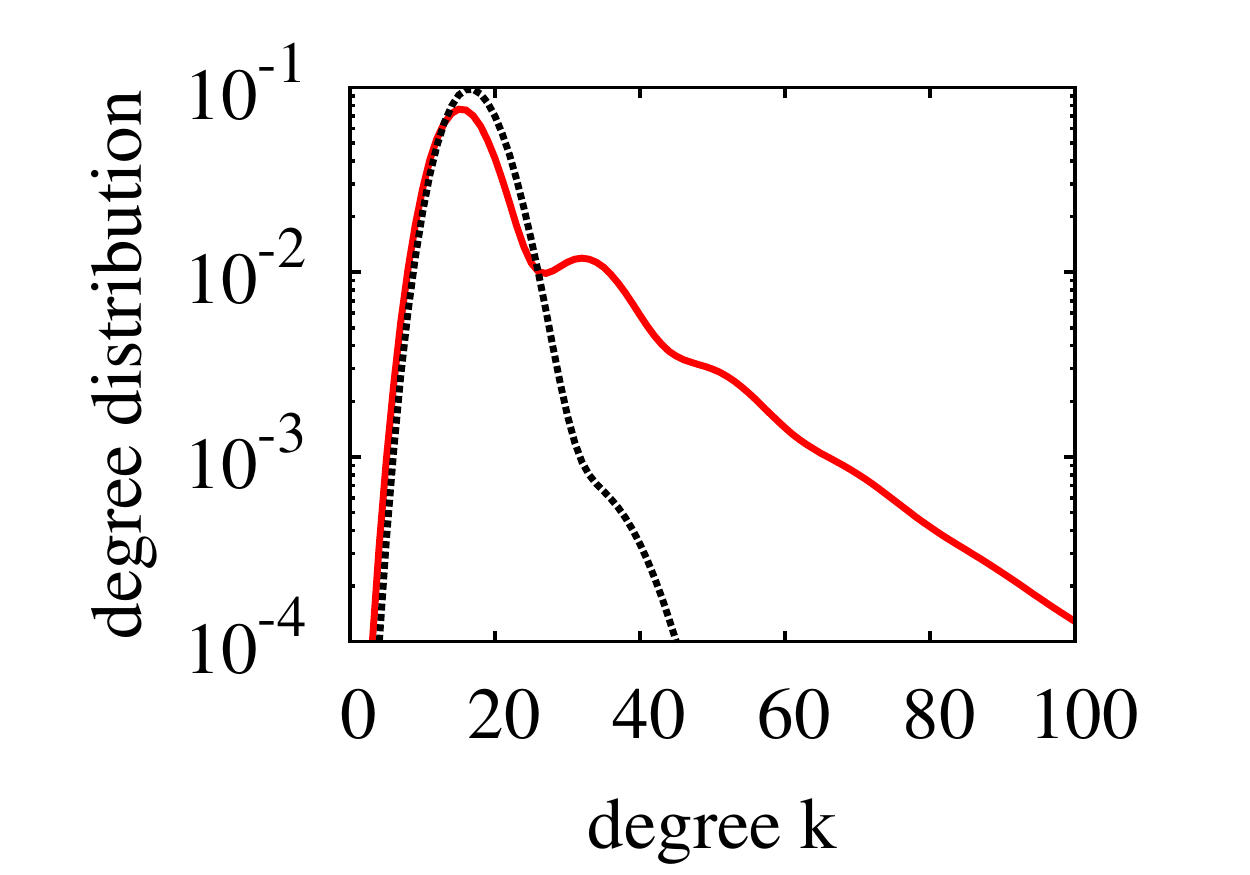}
\caption{(Color online) Degree distribution in the infinite network limit of the chosen topologies: (\ref{dis}) is shown by a solid shaded line while (\ref{dis2}) is shown by a dotted black line. Note the periodic local maxima corresponding to each $m$ value.}
\label{dis_fig}
\end{figure}

\subsection{Treatment of the analytical model}

\begin{figure*}[!htb]
  \centering
  \subfigure[]{\includegraphics[width=0.49\textwidth]{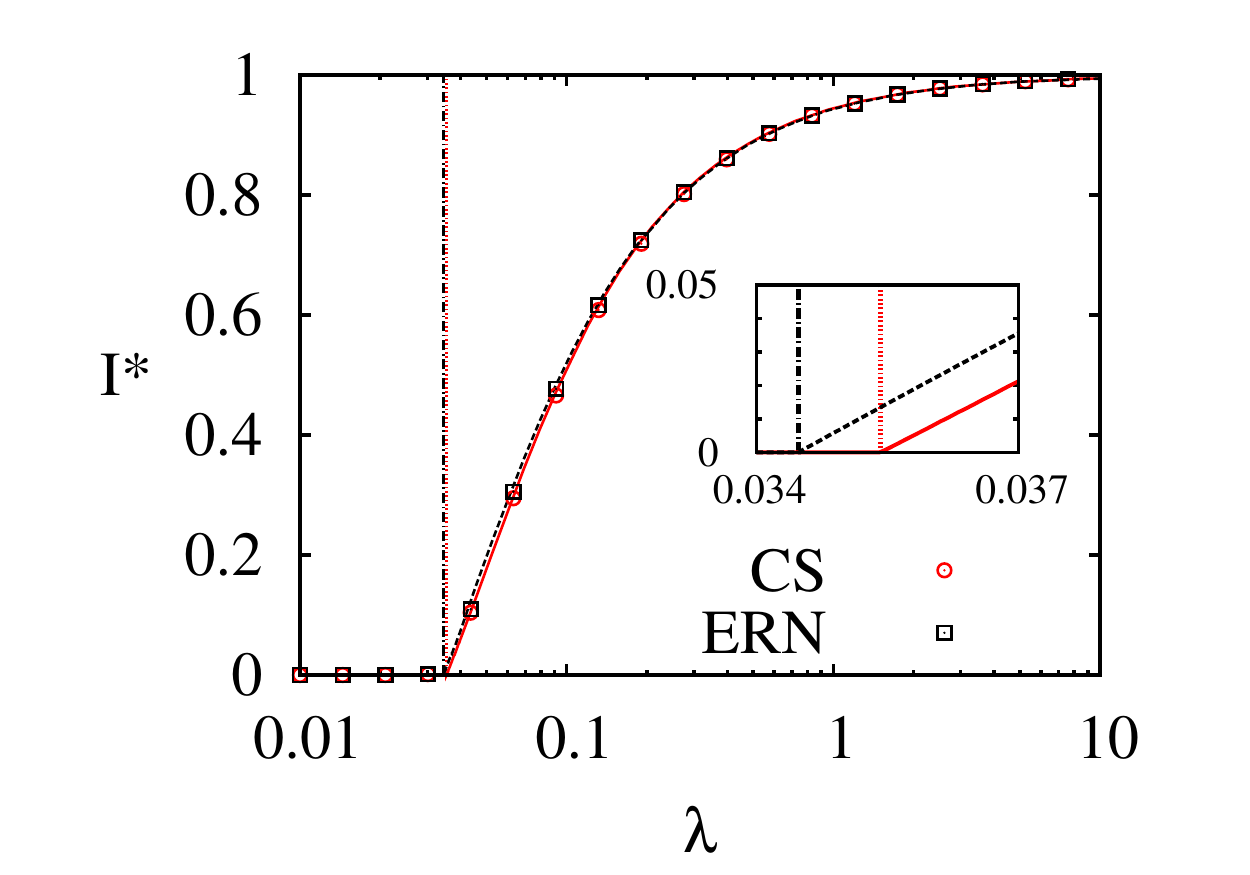} \label{stable}}            
  \subfigure[]{\includegraphics[width=0.49\textwidth]{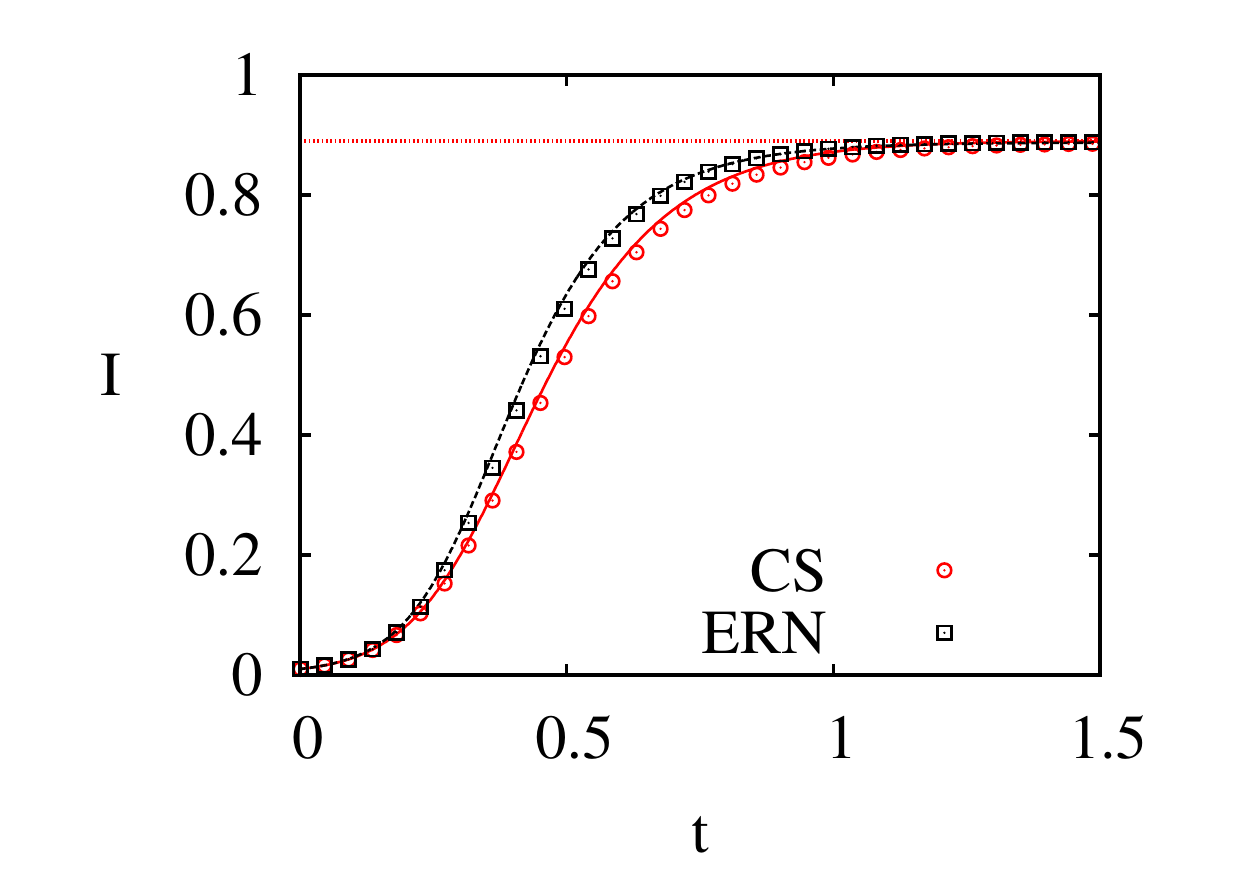} \label{evo_tempo}}
  \caption{(Color online) Comparisons of analytical and numerical results on a network defined by (\ref{topo}) using normalized dynamics ($t \rightarrow rt$ and $\lambda = \tau /r$). (a) analytical stable states (curves) and epidemic thresholds (vertical lines at $\lambda _c^{\textrm{CS}} = 3.54 \cdot 10^{-2}$ and $\lambda _c^{\textrm{ERN}} = 3.44 \cdot 10^{-2}$). (b) time evolution (curves) and analytical equilibrium (horizontal line) for $\lambda = 0.5$. On both figures, the results are shown in solid shade for the community structure (CS) and in dotted black line for the equivalent random network (ERN). Numerical results are presented by markers and are averaged over 20 000 networks of 25 000 nodes. The standard deviation is smaller than the marker size.}
  \label{results}
\end{figure*}

In order to highlight the difference between RN and CS, both types of networks will be studied analytically and numerically. The CS network will be compared with its equivalent random network (ERN): a network with exactly the same degree distribution, but with randomly connected nodes (zero degree correlation). Note that on our general model of community structure, the PGF for the degree distribution is simply generated by \cite{newman03a}:
\begin{equation}
G_0\left(P_1(1+(z-1)\epsilon)\right) \times K_0(z) \; .
\label{degree_PGF}
\end{equation}
To describe an ERN with this distribution, two simple options are available. Firstly, one can set $P_0^{\textrm{ERN}}(z) = z^2$ and $K_0^{\textrm{ERN}}(z) = 1$ with $\epsilon ^{\textrm{ERN}} = 1$ so that all cliques are of size two (i.e. regular links) and then choose the $g_m$ distribution equal to the initial degree distribution (\ref{degree_PGF}) of the CS network. Secondly, one can set $P_0^{\textrm{ERN}}(z) = z$ and $G_0^{\textrm{ERN}}(z) = z$ with any $\epsilon ^{\textrm{ERN}}$ so that all cliques are of size one (i.e. simple nodes) and then choose the $k_l$ distribution equal to the initial degree distribution (\ref{degree_PGF}). Both will be used in what follows.

The time evolution of the analytical system is obtained from an integration based on a 4th order Runge-Kutta algorithm with adaptive time steps. The initial condition $I(0)$ is uniformly distributed among the nodes. That is, $S_{m,l}(0) = g_mk_l\left(1-I(0)\right)$ for all $\{ m,l \}$, while $\{ C_{n,i}(0)\}$ are given by a simple Bernoulli trial:
\begin{equation}
C_{n,i}(0) = p_n \binom{n}{i}\left[I(0)\right]^i\left[1-I(0)\right]^{n-i} \; .
\end{equation}

\subsection{Numerical model}
To perform MC simulations of the model, we have generated networks with the structure presented in section \ref{section:CS} via the following numerical algorithm:
\begin{enumerate}
\item[i.]  generate a sequence $\{ m_i \}$ of length $N$ subjected to distribution $\{ g_m\}$;
\item[ii.]  generate a sequence $\{ n_j \}$ subjected to distribution $\{ p_n\}$ until $\sum _j n_j = \sum _i m_i$;
\item[iii.] for each $i$, produce $m_i$ individuals tagged as $i$;
\item[iv.] for each $j$, produce $n_j$ groups tagged as $j$;
\item[v.] randomly assign each individual to a group;
\item[vi.] for each $i$, list every $i$ assigned to the $n_j$ groups and link them to one another with probability $\epsilon$.          
\item[vii.]   generate a sequence $\{ l_s \}$ of length $N$ subjected to the distribution $\{ k_l\}$ under condition that $\sum _s l_s$ is even;
\item[viii.] for each $s$, produce $l_s$ stubs tagged as $s$;
\item[ix.] randomly link all stubs in pairs. 
\end{enumerate}
The final ensemble of links presents a topology as shown in Fig. \ref{schema} with a degree distribution generated by (\ref{degree_PGF}); where nodes are highly clustered, but the clique concept itself is invisible. Each and every network generated by this procedure is accepted and kept in the results, as they are part of the canonical ensemble considered by the mean-field approach of the formalism. For every generated network, a fraction $I(0)$ of individuals are randomly chosen to be initially infectious and the dynamics is then simulated in a discrete time propagation simulation valid for a time step $\Delta t \rightarrow 0$ (we choose $\Delta t$ such that $\tau\Delta t$ and $r\Delta t$ are lesser than $10^{-3}$):
\begin{enumerate} 
\item[i.]  at each $\Delta t$, every susceptible neighbor of every infectious individual is infected with probability $\tau \Delta t$;     
\item[ii.] at each $\Delta t$ every infectious individual recovers with probability $r \Delta t$.     
\end{enumerate}
Finally, for each constructed network, the final degree distribution is used to generate an ERN for comparison.

\begin{figure*}[!htb]
  \centering
  \subfigure[]{\includegraphics[width=0.49\textwidth]{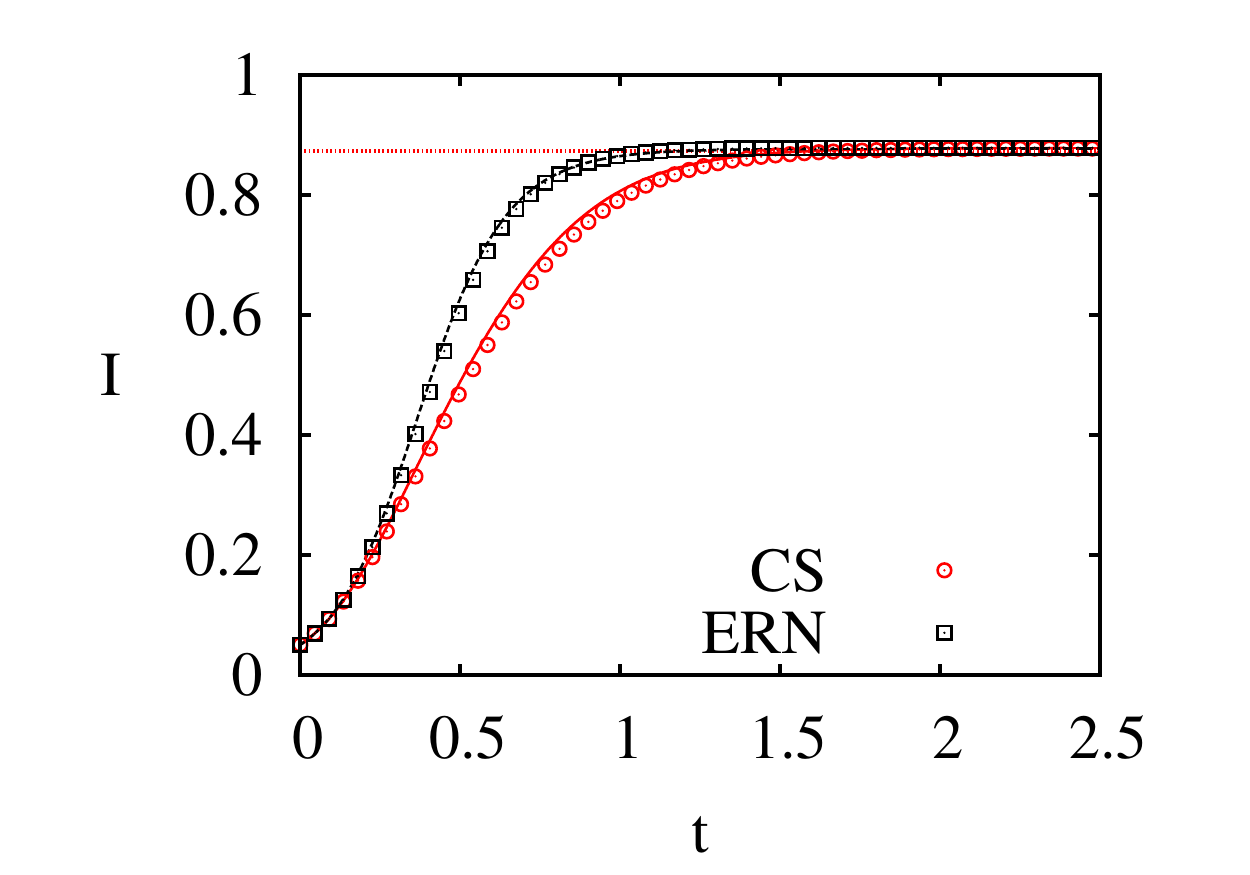}\label{evo2}}
  \subfigure[]{\includegraphics[width=0.49\textwidth]{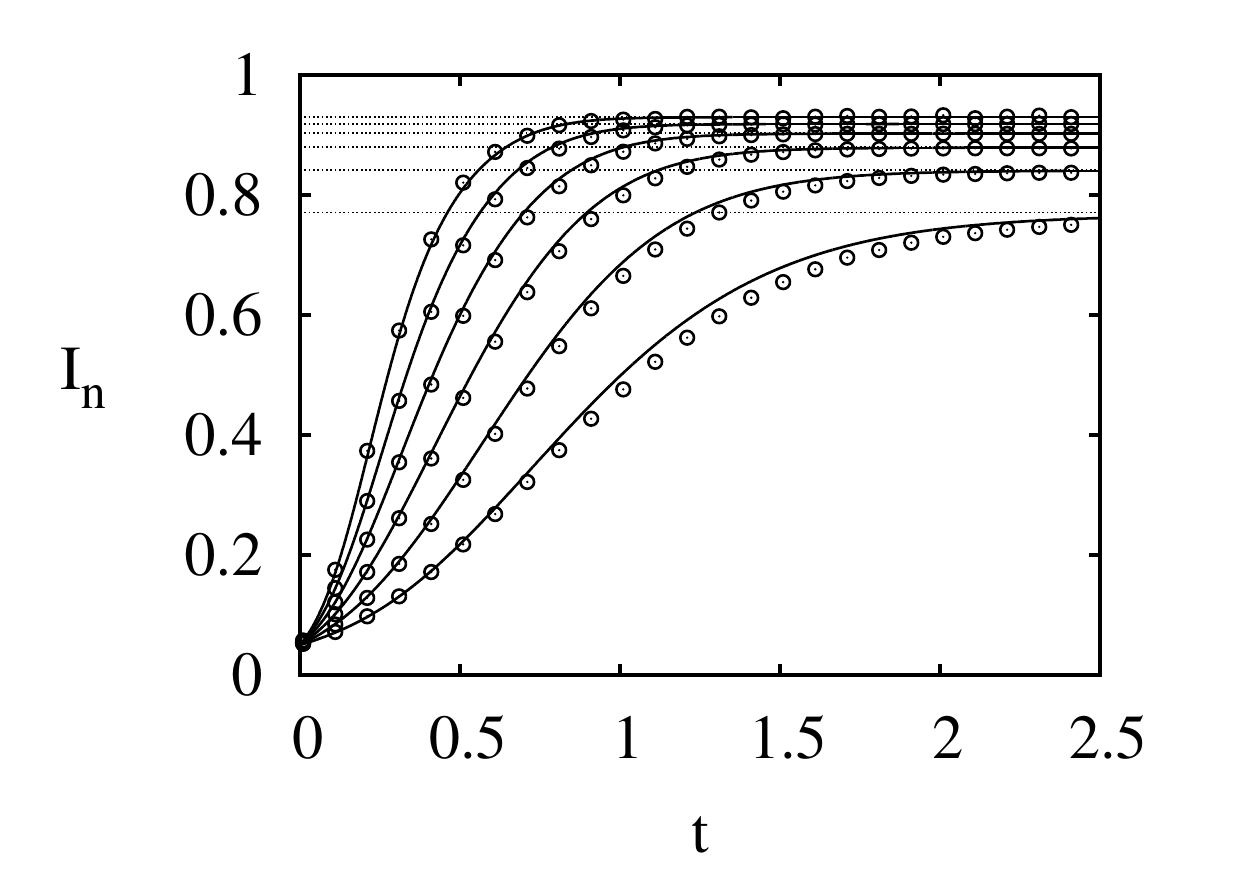}\label{all_clics}}
  \caption{(Color online) Comparison between analytical and numerical results on a network with general community structure defined by (\ref{topo2}) for a SIS model of propagation dynamics of parameter $\lambda = \tau /r = 0.5$ under normalized time $t \rightarrow rt$. (a) time evolution of the global state (community structure in solid shade and equivalent random network in dotted black) and (b) time evolution for cliques of size 10, 15, 20, 25, 30 and 35 (lowest to highest curves). All numerical results are obtained via MC simulations on over $20 000$ networks of $25 000$ nodes and are presented by their mean value. Analytical predictions for the stables states are shown in horizontal dotted lines in both figures. Note that the deviation from the predictions is bigger for the smallest cliques than for the larger ones. This is a consequence of the mean-field description which is more accurate for large systems (or, in this case, subsystems) for which standard deviations are of lesser relative importance.}
  \label{topo2_res}
\end{figure*}

\subsection{Results on Newman's topology}

The first topology chosen to test the formalism is the special model presented in \cite{newman03a}, which does not allow random links and is thus obtained by setting $K_0(z) = 1$ (i.e. all links are shared within a clique). We will then use $\epsilon = 0.8$, a power-law distribution for the numbers of cliques per individual and a Poisson distribution for the numbers of individuals per clique:
\begin{equation}
g_m \propto m^{-1}e^{-m/1.2} \quad ; \quad  p_n \propto \frac{20^n}{n!}e^{-20} \; .
\label{topo}
\end{equation}
This topology results in a degree distribution generated by the following function:
\begin{equation}
G_0\textbf{(}P_1 \textbf{(} 1+(z-1)\epsilon\textbf{)}\textbf{)} = \frac{\ln(1-e^{20(\epsilon z - \epsilon)}e^{-5/6})}{\ln(1-e^{-5/6})} \; .
\label{dis}
\end{equation}
This heterogenous distribution is shown in Fig. \ref{dis_fig}. To follow the propagation dynamics on an ERN, we use the first of the two options previously presented: all cliques are of size two with $\epsilon ^{\textrm{ERN}} =1$ and a distribution $\{ g_m \}$ equivalent to (\ref{dis}).

Our results on this topology, Fig. \ref{results}, confirm that our formalism is indeed capable of following the time evolution of the network in structured and random topologies. Furthermore, both our numerical and analytical results support the conclusions of \cite{huang07,hiebeler06,watts05} as will be discussed below.

Firstly, as evident in Fig. \ref{stable}, the community structure does not significantly change the stable state of the system. This conclusion is only valid when the giant components of CS and of the ERN have approximately the same size and under condition that the network is well connected. In physical terms, this means that the coupling must be sufficiently high between the subsystems, relative to the strength of the interaction (i.e. $\lambda$). If this condition is not fully met, subsets of the canonical distribution of configurations (i.e. ones with higher number of independent cliques) will have stable states under the predicted value and will decrease the mean value. The reduction of the giant component was already explained in \cite{newman03a}. This effect is visible in both analytical and numerical results of Fig. \ref{stable} for lower infection rate and eventually leads to a higher epidemic threshold for networks with community structure. 

This particular property seems to contradict a major conclusion of \cite{newman03a}, yet it is important to take into account that the conclusion that clustering lowers the epidemic threshold was made on networks featuring different degree distributions (see \cite{kiss08} for a complete discussion) and featuring degree correlation (see \cite{miller09} for an analysis of correlation and clustering effects). Our results show that, given an \emph{identical} degree distribution and zero degree correlation, the random networks will have a lower epidemic threshold than a network featuring community structure. This conclusion is intuitive because links shared in community have a higher probability of being ``wasted'' (i.e. of leading to another infectious node) than a random link, independently of the transmissibility. The mechanism behind this phenomenon is simple: there is a higher probability that neighbors of a new infectious individual will also be infectious if these individuals are connected in groups. This leads to a lower mean epidemic size for low infection rate and to the observed higher epidemic threshold. Note that, within the community structure effects observed here, the individual effects of clustering and degree correlation can not be separated. The demonstration given in Appendix shows that, for networks with zero degree correlation, our model always predicts a higher epidemic threshold for networks with clustering than for equivalent random networks. However, it should be emphasized that correlation effects alone have been shown to lower the percolation threshold \cite{gleeson10}. As similar effects can take place on networks with community structure, our conclusion is not directly generalizable to networks with non-zero degree correlation.
 
Secondly, as seen in Fig. \ref{evo_tempo}, the community structure increases the relaxation time of the system; i.e. it slows the disease propagation towards the equilibrium. This phenomenon is also explained by the higher number of wasted links on a community structure than on the equivalent random network. These links are very frequent in social networks because of community structure where ``the friend of my friend is also my friend''. When counting new possible infections on networks with exactly the same degree distribution, the number of second neighbors will be higher in a random network than on a community structure, because the neighbors of my neighbor may have already been counted as my neighbor in the CS network. This results in a slower propagation and a typically higher epidemic threshold.

Finally, note that the shift observed in the epidemic threshold is not always as small as seen on Fig. \ref{stable}. For example, a topology with $G'(1) \simeq 0.365$ and $\nu = 5$ yields $\lambda _c ^{\textrm{CS}} = 5/4 \cdot \lambda _c^{\textrm{ERN}}$. This particular case was verified by MC simulations.

\subsection{Results on a general topology}

As a second test to our formalism, we use $\epsilon =0.8$ and the following distributions:
\begin{eqnarray}
g_m \propto \frac{e^{-4m}}{m}; \quad p_n \propto \frac{20^n}{n!}e^{-20}; \quad k_l \propto \frac{e^{-l}}{l}
\label{topo2}
\end{eqnarray}
which result in the second degree distribution shown in Fig. \ref{dis_fig} and generated by:
\begin{equation}
\frac{\ln(1-e^{20(\epsilon z - \epsilon)}e^{-5/6})}{\ln(1-e^{-5/6})}\frac{\ln(1-ze^{-1})}{\ln(1-e^{-1})} \; .
\label{dis2}
\end{equation}
In this case, the ERN are obtained by using cliques of size one and fitting the degree distribution with the random links generated by $K_0(z)$. The results obtained on this second topology are presented in Fig. \ref{topo2_res}. They not only confirm the quality of our treatment, but also earlier conclusions. The propagation slow-down is stronger in the time evolution featured in Fig. \ref{evo2} than in the case observed in Fig. \ref{evo_tempo}, because the topology used produces a much higher proportion of intra-clique links for a given individual, and consequently, a higher fraction of wasted links. It is believed that this effect could be studied using percolation theory with a quantification of CS, such as the modularity concept introduced by Newman and Girvan in \cite{newman04a}.

\section{Conclusion \label{section:conclu}}
What may well be the single most important contribution of this paper is the philosophy upon which the formalism is based. An effective dynamical description of complex networks can be obtained by a mean-field approach using a compartmentalisation of both the networks' elements (e.g. individuals or nodes) and of their recurrent topological patterns (e.g. cliques or substructures) in classes of homogeneous state and behavior. It has been shown that a particular topology, the community structure, can be solved with this method. Furthermore, the approach can also describe random topology in the limit of the most elementary patterns possible. Hence, it is reasonable to assert that other complex topologies may be treated in a similar manner.

More precisely, our analytical results confirm previous numerical simulations on the effects of community structure in propagation dynamics: in comparison to equivalent random networks, the structured systems feature longer relaxation times (i.e. slower propagation) and generally higher epidemic thresholds. 

An especially interesting avenue to explore would be to direct the formalism towards more epidemiologically oriented applications with a generalization to other propagation model (see for example \cite{ball10}). Furthermore, in an epidemic context, taking the topology of social network into account allows precise emulation of real intervention scenarios which are often based on groups of individuals (e.g. school closings and vaccination of public health workers both correspond to interventions on given cliques).

Other applications of our formalism are possible in various models of dynamics and topologies. Of particular interest is the application of our formalism to dynamical networks (e.g. \cite{gross09, marceau10}). This may help in gaining insights on the emergence and the stability of social structure.

\begin{acknowledgments}
The research team is grateful to CIHR (LHD, PAN and AA), NSERC (VM and LJD) and FQRNT (LJD) for financial support.
\end{acknowledgments}

\appendix*

\section{Community structure, without degree correlation, raises the epidemic threshold \label{section:proof}}
This paper has shown that our model can describe propagation phenomena on network with community structure as well as network with random topology. Using the analytic solution for the epidemic threshold on Newman's topology, it is possible to show that, given two networks with identical degree distributions and zero degree correlation, but where one is completely random while the other features community structure (and therefore clustering), the latter will have a higher epidemic threshold.

First of all, degree correlation refers to situations where, given a random link in the network, the knowledge of the excess degree of one of its nodes influences the probability distribution for the excess degree of the other. For Newman's model, it was shown in \cite{newman03b} that the probability $e_{jk}$ that a given link joins two nodes of excess degree $j$ and $k$ can be calculated as follows. We first write:
\begin{equation}
e_{jk} = \frac{1}{N}\sum _n p_n n(n-1)P(j,k\vert n) \; ,
\label{ejk}
\end{equation}
where $n(n-1)$ is the number of potential degrees in a clique of size $n$, $N$ is a normalization factor corresponding to the total number of potential links in the network and $P(j,k\vert n)$ is the probability that a link within a clique of size $n$ joins two nodes of excess degree $j$ and $k$. This probability can be calculated by separating $j$ in $j_{\textrm{in}}$ and $j_{\textrm{out}}$, respectively the excess links shared within and outside of the considered clique, and doing the same for $k$. We can now write:
\begin{align}
P(j,k\vert n) = & \sum _{j_\textrm{in}} \binom{n-2}{j_\textrm{in}} \epsilon ^{j_\textrm{in}}(1-\epsilon)^{n-2-j_\textrm{in}}P(j_{\textrm{out}}) \nonumber \\
 & \!\!\!\!\!\!\!\! \!\!\!\!   + \sum _{k_\textrm{in}} \binom{n-2}{k_\textrm{in}} \epsilon ^{k_\textrm{in}}(1-\epsilon)^{n-2-k_\textrm{in}}P(k_{\textrm{out}}) \; ,
 \label{Pjkn}
\end{align}
where $P(j_{\textrm{out}})$ and $P(k_{\textrm{out}})$ are the probabilities that the nodes have $j_{\textrm{out}}$ and $k_{\textrm{out}}$ links outside the clique of size $n$. These two probabilities are simply generated by the PGFs composition $G_1\left(P_1(1+(z-1)\epsilon)\right)$. Now, because both $k$ and $j$ must be calculated with one clique in common where they both have $n-2$ potential excess neighbors, we can write the set of $\{ e_{jk} \}$ in terms of the following PGF:
\begin{align}
\sum _{jk} e_{jk}x^jy^k = & P_2\left((1+(x-1)\epsilon)(1+(y-1)\epsilon)\right) \nonumber \\ 
& \!\!\!\!\!\!\!\!\!\!\!\!\!\!\!\!\!\!\!\!\!\!\!\! \times G_1\left(P_1(1\! +\! (x\! -\! 1)\epsilon)\right)G_1\left(P_1(1\! +\! (y\! -\! 1)\epsilon)\right) \; ,
\label{ejk_PGF_CS}
\end{align}
where $P_2(z) \equiv \left[P_0''(1)\right]^{-1}\sum _n n(n-1)z^{n-2}$. For a random network, it is easily obtained that $e_{jk}$ is simply the product of the two independent probabilities of having nodes of excess degree $j$ and $k$. Thus, by differentiating the degree distribution PGF (\ref{degree_PGF}) to obtain the excess degree distribution, we find:
\begin{align}
\sum _{jk} e^{ERN}_{jk}x^jy^k = & P_2\left(1\! +\! (x\! -\! 1)\epsilon\right)G_1\left(P_1(1\! +\! (x\! -\! 1)\epsilon)\right) \nonumber \\\ 
& \!\!\!\!\!\!\!\!\!\!\!\!\!\!\!\!\!\!\!\!\!\!\!\! \times P_2\left(1\! +\! (y\! -\! 1)\epsilon\right)G_1\left(P_1(1\! +\! (y\! -\! 1)\epsilon)\right) \; .
\label{ejk_PGF_RG}
\end{align}
For expressions (\ref{ejk_PGF_CS}) and (\ref{ejk_PGF_RG}) to be equivalent, the following condition must be satisfied:
\begin{align}
& P_2\left((1+(x-1)\epsilon)(1+(y-1)\epsilon)\right) \nonumber \\
& \qquad\qquad = P_2\left(1 +(x - 1)\epsilon \right)P_2\left(1 + (y - 1)\epsilon \right) \; .
\label{p2_cond}
\end{align}
We want to compare two networks sharing exactly the same degree distribution and degree correlation. Equation (\ref{p2_cond}) gives us the condition for which two networks with identical degree distributions, one featuring community structure and the other random topology, will have the same degree correlation. It is easy to conclude that the distribution of individuals per clique, in order to respect Eq. (\ref{p2_cond}), can only be given by:
\begin{equation}
p_n = \delta _{n,\nu}
\end{equation}
where $\nu$ is an arbitrary positive integer. In other words, all structures must be the same size. This limitation comes from the way we construct our random networks. Because by simply matching degrees generated from a given distribution, the knowledge of one neighbor's degree does not give any information concerning the other neighbor's degree. Note that $G_0(z)$ and $\epsilon$ are totally free, so that the heterogeneity of the degree distribution is not entirely compromised.

We will now compare two networks with zero degree correlations. The first is random with $p^{\textrm{ERN}}_n = \delta _{n,2}$ and $\epsilon ^{\textrm{ERN}} = 1$ while the other exhibits community structure with $p^{\textrm{CS}}_n = \delta _{n, \nu}$ with $\nu > 2$ and $\epsilon ^{\textrm{CS}} \equiv \epsilon \in [0,1]$. The two networks have exactly the same degree distribution, which means that $G_0^{\textrm{ERN}}(z) = G_0^{\textrm{CS}}\left(P^{\textrm{CS}}(1+(z-1)\epsilon)\right)$. Using Eq. (\ref{thresh_poly}), we can easily write the epidemic threshold for the random network:
\begin{equation}
\lambda _c ^{\textrm{ERN}} = \dfrac{1}{\frac{d}{dz}G_1^{\textrm{ERN}}(1)} \equiv \dfrac{1}{\epsilon\left[(\nu -2) + \mu _1(\nu -1)\right]} \; ,
\label{tau_c_ann}
\end{equation}
where the last expression uses the PGFs of the structured network in which $\mu _1 = \frac{d}{dz}G_1^{\textrm{CS}}(1)$ is the mean number of excess cliques per individual. We will now insert expression (\ref{tau_c_ann}) in the epidemic threshold condition (\ref{thresh_cond}) of the network with community structure. Because all terms in the polynomial are positive, we expect to find an expression greater than unity if (\ref{tau_c_ann}) is higher than the threshold for CS, equal to one if the threshold remains the same or lesser than unity if the threshold for the ERN is actually lower than that for CS. To prove the latter case, for arbitrary $\nu$, $\epsilon$ and $\{ g_m \}$, we simply demonstrate the following inequality written from (\ref{thresh_cond}) using (\ref{tau_c_ann}):
\begin{equation}
\sum _{i=1}^{\nu -1} \frac{\mu _1 (\nu -1)!}{(\nu - i -1)!}\left[(\nu - 2) + \mu _1 (\nu -1)\right]^{-i} < 1 \; .
\label{ineq}
\end{equation}
Further, it can be shown that the derivative of (\ref{ineq}) in $\mu _1$ is always positive. This provides us with an upper bound for (\ref{ineq}) in the limit $\mu _1 \rightarrow \infty$. Using l'H\^opital's rule, we thus find:
\begin{equation}
\lim _{\mu _1 \rightarrow \infty}  \sum _{i=1}^{\nu -1} \!\frac{\mu _1 (\nu -1)!}{(\nu - i -1)!}\left[(\nu - 2)\! +\! \mu _1 (\nu -1)\right]^{-i} \! =\! 1 \; .
\end{equation}
This indicates that the two networks with zero degree correlation, one featuring community structure and one an equivalent random network, will have the same threshold in the limit of infinite mean number of excess cliques per individual or if $\nu = 2$. Otherwise, because the derivative of the polynomial in $\mu _1$ was shown to be positive, finite $\mu _1$ and $\nu > 2$ imply a higher threshold for the structured network.


\begin{thebibliography}{43}
\expandafter\ifx\csname natexlab\endcsname\relax\def\natexlab#1{#1}\fi
\expandafter\ifx\csname bibnamefont\endcsname\relax
  \def\bibnamefont#1{#1}\fi
\expandafter\ifx\csname bibfnamefont\endcsname\relax
  \def\bibfnamefont#1{#1}\fi
\expandafter\ifx\csname citenamefont\endcsname\relax
  \def\citenamefont#1{#1}\fi
\expandafter\ifx\csname url\endcsname\relax
  \def\url#1{\texttt{#1}}\fi
\expandafter\ifx\csname urlprefix\endcsname\relax\def\urlprefix{URL }\fi
\providecommand{\bibinfo}[2]{#2}
\providecommand{\eprint}[2][]{\url{#2}}

\bibitem[{\citenamefont{Barrat et~al.}(2008)\citenamefont{Barrat, Barth\'elemy,
  and Vespignani}}]{barrat08}
\bibinfo{author}{\bibfnamefont{A.}~\bibnamefont{Barrat}},
  \bibinfo{author}{\bibfnamefont{M.}~\bibnamefont{Barth\'elemy}},
  \bibnamefont{and}
  \bibinfo{author}{\bibfnamefont{A.}~\bibnamefont{Vespignani}},
  \emph{\bibinfo{title}{Dynamical Processes on Complex Networks}}
  (\bibinfo{publisher}{Cambridge University Press}, \bibinfo{year}{2008}).

\bibitem[{\citenamefont{Anderson and May}(1991)}]{anderson91}
\bibinfo{author}{\bibfnamefont{R.~M.} \bibnamefont{Anderson}} \bibnamefont{and}
  \bibinfo{author}{\bibfnamefont{R.~M.} \bibnamefont{May}},
  \emph{\bibinfo{title}{Infectious Disease of Humans: Dynamics and Control}}
  (\bibinfo{publisher}{Oxford University Press}, \bibinfo{year}{1991}).

\bibitem[{\citenamefont{Newman et~al.}(2001)\citenamefont{Newman, Strogatz, and
  Watts}}]{newman01}
\bibinfo{author}{\bibfnamefont{M.~E.~J.} \bibnamefont{Newman}},
  \bibinfo{author}{\bibfnamefont{S.}~\bibnamefont{Strogatz}}, \bibnamefont{and}
  \bibinfo{author}{\bibfnamefont{D.}~\bibnamefont{Watts}},
  \bibinfo{journal}{Phys. Rev. E} \textbf{\bibinfo{volume}{64}},
  \bibinfo{pages}{026118} (\bibinfo{year}{2001}).

\bibitem[{\citenamefont{Newman}(2002)}]{newman02}
\bibinfo{author}{\bibfnamefont{M.~E.~J.} \bibnamefont{Newman}},
  \bibinfo{journal}{Phys. Rev. E} \textbf{\bibinfo{volume}{66}},
  \bibinfo{pages}{016128} (\bibinfo{year}{2002}).

\bibitem[{\citenamefont{Allard et~al.}(2009)\citenamefont{Allard, No\"el,
  Dub{\'e}, and Pourbohloul}}]{allard09}
\bibinfo{author}{\bibfnamefont{A.}~\bibnamefont{Allard}},
  \bibinfo{author}{\bibfnamefont{P.-A.} \bibnamefont{No\"el}},
  \bibinfo{author}{\bibfnamefont{L.~J.} \bibnamefont{Dub{\'e}}},
  \bibnamefont{and}
  \bibinfo{author}{\bibfnamefont{B.}~\bibnamefont{Pourbohloul}},
  \bibinfo{journal}{Phys. Rev. E} \textbf{\bibinfo{volume}{79}},
  \bibinfo{pages}{036113} (\bibinfo{year}{2009}).

\bibitem[{\citenamefont{No\"{e}l et~al.}(2009)\citenamefont{No\"{e}l, Davoudi,
  Brunham, Dub\'{e}, and Pourbohloul}}]{noel09}
\bibinfo{author}{\bibfnamefont{P.-A.} \bibnamefont{No\"{e}l}},
  \bibinfo{author}{\bibfnamefont{B.}~\bibnamefont{Davoudi}},
  \bibinfo{author}{\bibfnamefont{R.~C.} \bibnamefont{Brunham}},
  \bibinfo{author}{\bibfnamefont{L.~J.} \bibnamefont{Dub\'{e}}},
  \bibnamefont{and}
  \bibinfo{author}{\bibfnamefont{B.}~\bibnamefont{Pourbohloul}},
  \bibinfo{journal}{Phys. Rev. E} \textbf{\bibinfo{volume}{79}},
  \bibinfo{pages}{026101} (\bibinfo{year}{2009}).

\bibitem[{\citenamefont{Volz}(2008)}]{volz08}
\bibinfo{author}{\bibfnamefont{E.}~\bibnamefont{Volz}}, \bibinfo{journal}{j.
  Math. Biol.} \textbf{\bibinfo{volume}{56}}, \bibinfo{pages}{293}
  (\bibinfo{year}{2008}).

\bibitem[{\citenamefont{Marder}(2007)}]{marder07}
\bibinfo{author}{\bibfnamefont{M.}~\bibnamefont{Marder}},
  \bibinfo{journal}{Phys. Rev. E} \textbf{\bibinfo{volume}{75}},
  \bibinfo{pages}{066103} (\bibinfo{year}{2007}).

\bibitem[{\citenamefont{Pautasso and Jeger}(2008)}]{pautasso08}
\bibinfo{author}{\bibfnamefont{M.}~\bibnamefont{Pautasso}} \bibnamefont{and}
  \bibinfo{author}{\bibfnamefont{M.~J.} \bibnamefont{Jeger}},
  \bibinfo{journal}{Ecol. Compl. 5} \textbf{\bibinfo{volume}{5}},
  \bibinfo{pages}{1} (\bibinfo{year}{2008}).

\bibitem[{\citenamefont{May}(2006)}]{may06}
\bibinfo{author}{\bibfnamefont{R.~M.} \bibnamefont{May}},
  \bibinfo{journal}{Trends Ecol. Evol.} \textbf{\bibinfo{volume}{21}},
  \bibinfo{pages}{394} (\bibinfo{year}{2006}).

\bibitem[{\citenamefont{Keeling}(2005)}]{keeling05}
\bibinfo{author}{\bibfnamefont{M.}~\bibnamefont{Keeling}},
  \bibinfo{journal}{Theor. Popul. Biol.} \textbf{\bibinfo{volume}{67}},
  \bibinfo{pages}{1} (\bibinfo{year}{2005}).

\bibitem[{\citenamefont{Keeling and Eames}(2005)}]{keeling05b}
\bibinfo{author}{\bibfnamefont{M.~J.} \bibnamefont{Keeling}} \bibnamefont{and}
  \bibinfo{author}{\bibfnamefont{K.~T.~D.} \bibnamefont{Eames}},
  \bibinfo{journal}{J. R. Soc. Interface} \textbf{\bibinfo{volume}{2}},
  \bibinfo{pages}{295} (\bibinfo{year}{2005}).

\bibitem[{\citenamefont{Shirley and Rushton}(2005)}]{shirley05}
\bibinfo{author}{\bibfnamefont{M.~D.} \bibnamefont{Shirley}} \bibnamefont{and}
  \bibinfo{author}{\bibfnamefont{S.~P.} \bibnamefont{Rushton}},
  \bibinfo{journal}{Ecol. Compl.} \textbf{\bibinfo{volume}{2}},
  \bibinfo{pages}{287} (\bibinfo{year}{2005}).

\bibitem[{\citenamefont{Miller}(2009)}]{miller09}
\bibinfo{author}{\bibfnamefont{J.~C.} \bibnamefont{Miller}},
  \bibinfo{journal}{Phys. Rev. E} \textbf{\bibinfo{volume}{80}},
  \bibinfo{pages}{020901} (\bibinfo{year}{2009}).

\bibitem[{\citenamefont{Gleeson}(2009)}]{gleeson09}
\bibinfo{author}{\bibfnamefont{J.~P.} \bibnamefont{Gleeson}},
  \bibinfo{journal}{Phys. Rev. E} \textbf{\bibinfo{volume}{80}},
  \bibinfo{pages}{036107} (\bibinfo{year}{2009}).

\bibitem[{\citenamefont{Newman}(2009)}]{newman09}
\bibinfo{author}{\bibfnamefont{M.~E.~J.} \bibnamefont{Newman}},
  \bibinfo{journal}{Phys. Rev. Lett.} \textbf{\bibinfo{volume}{103}},
  \bibinfo{pages}{058701} (\bibinfo{year}{2009}).

\bibitem[{\citenamefont{Ferrari et~al.}(2006)\citenamefont{Ferrari, Bansal,
  Meyers, and Bj{\o}rnstad}}]{ferrari06}
\bibinfo{author}{\bibfnamefont{M.~J.} \bibnamefont{Ferrari}},
  \bibinfo{author}{\bibfnamefont{S.}~\bibnamefont{Bansal}},
  \bibinfo{author}{\bibfnamefont{L.~A.} \bibnamefont{Meyers}},
  \bibnamefont{and} \bibinfo{author}{\bibfnamefont{O.~N.}
  \bibnamefont{Bj{\o}rnstad}}, \bibinfo{journal}{Proc. R. Soc. B}
  \textbf{\bibinfo{volume}{273}}, \bibinfo{pages}{2743} (\bibinfo{year}{2006}).

\bibitem[{\citenamefont{Newman}(2003)}]{newman03a}
\bibinfo{author}{\bibfnamefont{M.~E.~J.} \bibnamefont{Newman}},
  \bibinfo{journal}{Phys. Rev. E} \textbf{\bibinfo{volume}{68}},
  \bibinfo{pages}{026121} (\bibinfo{year}{2003}).

\bibitem[{\citenamefont{Newman and Park}(2003)}]{newman03b}
\bibinfo{author}{\bibfnamefont{M.~E.~J.} \bibnamefont{Newman}}
  \bibnamefont{and} \bibinfo{author}{\bibfnamefont{J.}~\bibnamefont{Park}},
  \bibinfo{journal}{Phys. Rev. E} \textbf{\bibinfo{volume}{68}},
  \bibinfo{pages}{036122} (\bibinfo{year}{2003}).

\bibitem[{\citenamefont{Ravasz et~al.}(2002)\citenamefont{Ravasz, Somera,
  Mongru, Oltvai, and Barab\'{a}si}}]{ravasz02}
\bibinfo{author}{\bibfnamefont{E.}~\bibnamefont{Ravasz}},
  \bibinfo{author}{\bibfnamefont{A.~L.} \bibnamefont{Somera}},
  \bibinfo{author}{\bibfnamefont{D.~A.} \bibnamefont{Mongru}},
  \bibinfo{author}{\bibfnamefont{Z.~N.} \bibnamefont{Oltvai}},
  \bibnamefont{and} \bibinfo{author}{\bibfnamefont{A.-L.}
  \bibnamefont{Barab\'{a}si}}, \bibinfo{journal}{Science}
  \textbf{\bibinfo{volume}{297}}, \bibinfo{pages}{1551} (\bibinfo{year}{2002}).

\bibitem[{\citenamefont{Spirin and Mirny}(2003)}]{spirin03}
\bibinfo{author}{\bibfnamefont{V.}~\bibnamefont{Spirin}} \bibnamefont{and}
  \bibinfo{author}{\bibfnamefont{L.~A.} \bibnamefont{Mirny}},
  \bibinfo{journal}{PNAS} \textbf{\bibinfo{volume}{100}},
  \bibinfo{pages}{12123} (\bibinfo{year}{2003}).

\bibitem[{\citenamefont{Onnela et~al.}(2003)\citenamefont{Onnela, Chakraborti,
  Kaski, Kert\'{e}sz, and Kanto}}]{onnela03}
\bibinfo{author}{\bibfnamefont{J.-P.} \bibnamefont{Onnela}},
  \bibinfo{author}{\bibfnamefont{A.}~\bibnamefont{Chakraborti}},
  \bibinfo{author}{\bibfnamefont{K.}~\bibnamefont{Kaski}},
  \bibinfo{author}{\bibfnamefont{J.}~\bibnamefont{Kert\'{e}sz}},
  \bibnamefont{and} \bibinfo{author}{\bibfnamefont{A.}~\bibnamefont{Kanto}},
  \bibinfo{journal}{Phys. Rev. E} \textbf{\bibinfo{volume}{68}},
  \bibinfo{pages}{056110} (\bibinfo{year}{2003}).

\bibitem[{\citenamefont{Heimo et~al.}(2007)\citenamefont{Heimo, Saram\"aki,
  Onnela, and Kaski}}]{heimo07}
\bibinfo{author}{\bibfnamefont{T.}~\bibnamefont{Heimo}},
  \bibinfo{author}{\bibfnamefont{J.}~\bibnamefont{Saram\"aki}},
  \bibinfo{author}{\bibfnamefont{J.-P.} \bibnamefont{Onnela}},
  \bibnamefont{and} \bibinfo{author}{\bibfnamefont{K.}~\bibnamefont{Kaski}},
  \bibinfo{journal}{Physica A} \textbf{\bibinfo{volume}{383}},
  \bibinfo{pages}{147} (\bibinfo{year}{2007}).

\bibitem[{\citenamefont{Palla et~al.}(2007)\citenamefont{Palla, Barab\'{a}si,
  and Vicsek}}]{palla07}
\bibinfo{author}{\bibfnamefont{G.}~\bibnamefont{Palla}},
  \bibinfo{author}{\bibfnamefont{A.-L.} \bibnamefont{Barab\'{a}si}},
  \bibnamefont{and} \bibinfo{author}{\bibfnamefont{T.}~\bibnamefont{Vicsek}},
  \bibinfo{journal}{Nature} \textbf{\bibinfo{volume}{446}},
  \bibinfo{pages}{664} (\bibinfo{year}{2007}).

\bibitem[{\citenamefont{Girvan and Newman}(2002)}]{girvan02}
\bibinfo{author}{\bibfnamefont{M.}~\bibnamefont{Girvan}} \bibnamefont{and}
  \bibinfo{author}{\bibfnamefont{M.~E.~J.} \bibnamefont{Newman}},
  \bibinfo{journal}{PNAS} \textbf{\bibinfo{volume}{99}}, \bibinfo{pages}{7821}
  (\bibinfo{year}{2002}).

\bibitem[{\citenamefont{Karrer et~al.}(2008)\citenamefont{Karrer, Levina, and
  Newman}}]{karrer08}
\bibinfo{author}{\bibfnamefont{B.}~\bibnamefont{Karrer}},
  \bibinfo{author}{\bibfnamefont{E.}~\bibnamefont{Levina}}, \bibnamefont{and}
  \bibinfo{author}{\bibfnamefont{M.~E.~J.} \bibnamefont{Newman}},
  \bibinfo{journal}{Phys. Rev. E} \textbf{\bibinfo{volume}{77}},
  \bibinfo{pages}{046119} (\bibinfo{year}{2008}).

\bibitem[{\citenamefont{Newman}(2006)}]{newman06}
\bibinfo{author}{\bibfnamefont{M.~E.~J.} \bibnamefont{Newman}},
  \bibinfo{journal}{PNAS} \textbf{\bibinfo{volume}{103}}, \bibinfo{pages}{8577}
  (\bibinfo{year}{2006}).

\bibitem[{\citenamefont{Pujol et~al.}(2006)\citenamefont{Pujol, B\'{e}jar, and
  Delgado}}]{pujol06}
\bibinfo{author}{\bibfnamefont{J.~M.} \bibnamefont{Pujol}},
  \bibinfo{author}{\bibfnamefont{J.}~\bibnamefont{B\'{e}jar}},
  \bibnamefont{and} \bibinfo{author}{\bibfnamefont{J.}~\bibnamefont{Delgado}},
  \bibinfo{journal}{Phys. Rev. E} \textbf{\bibinfo{volume}{74}},
  \bibinfo{pages}{016107} (\bibinfo{year}{2006}).

\bibitem[{\citenamefont{Newman and Girvan}(2004)}]{newman04a}
\bibinfo{author}{\bibfnamefont{M.~E.~J.} \bibnamefont{Newman}}
  \bibnamefont{and} \bibinfo{author}{\bibfnamefont{M.}~\bibnamefont{Girvan}},
  \bibinfo{journal}{Phys. Rev. E} \textbf{\bibinfo{volume}{69}},
  \bibinfo{pages}{026113} (\bibinfo{year}{2004}).

\bibitem[{\citenamefont{Newman}(2004)}]{newman04b}
\bibinfo{author}{\bibfnamefont{M.~E.~J.} \bibnamefont{Newman}},
  \bibinfo{journal}{Eur. Phys. J. B} \textbf{\bibinfo{volume}{38}},
  \bibinfo{pages}{321} (\bibinfo{year}{2004}).

\bibitem[{\citenamefont{Clauset et~al.}(2004)\citenamefont{Clauset, Newman, and
  Moore}}]{newman04c}
\bibinfo{author}{\bibfnamefont{A.}~\bibnamefont{Clauset}},
  \bibinfo{author}{\bibfnamefont{M.~E.~J.} \bibnamefont{Newman}},
  \bibnamefont{and} \bibinfo{author}{\bibfnamefont{C.}~\bibnamefont{Moore}},
  \bibinfo{journal}{Phys. Rev. E} \textbf{\bibinfo{volume}{70}},
  \bibinfo{pages}{066111} (\bibinfo{year}{2004}).

\bibitem[{\citenamefont{Huang and Li}(2007)}]{huang07}
\bibinfo{author}{\bibfnamefont{W.}~\bibnamefont{Huang}} \bibnamefont{and}
  \bibinfo{author}{\bibfnamefont{C.}~\bibnamefont{Li}}, \bibinfo{journal}{J.
  Stat. Mech.} p. \bibinfo{pages}{P01014} (\bibinfo{year}{2007}).

\bibitem[{\citenamefont{Ball}(1999)}]{ball99}
\bibinfo{author}{\bibfnamefont{F.}~\bibnamefont{Ball}}, \bibinfo{journal}{Math.
  Biosci.} \textbf{\bibinfo{volume}{156}}, \bibinfo{pages}{41}
  (\bibinfo{year}{1999}).

\bibitem[{\citenamefont{Ghoshal et~al.}(2004)\citenamefont{Ghoshal, Sander, and
  Sokolov}}]{ghoshal04}
\bibinfo{author}{\bibfnamefont{G.}~\bibnamefont{Ghoshal}},
  \bibinfo{author}{\bibfnamefont{L.}~\bibnamefont{Sander}}, \bibnamefont{and}
  \bibinfo{author}{\bibfnamefont{I.}~\bibnamefont{Sokolov}},
  \bibinfo{journal}{Math. Biosci.} \textbf{\bibinfo{volume}{190}},
  \bibinfo{pages}{71} (\bibinfo{year}{2004}).

\bibitem[{\citenamefont{Hiebeler}(2006)}]{hiebeler06}
\bibinfo{author}{\bibfnamefont{D.}~\bibnamefont{Hiebeler}},
  \bibinfo{journal}{Bull. Math. Biol.} \textbf{\bibinfo{volume}{68}},
  \bibinfo{pages}{1345} (\bibinfo{year}{2006}).

\bibitem[{\citenamefont{Pastor-Satorras and Vespignani}(2001)}]{vespignani01}
\bibinfo{author}{\bibfnamefont{R.}~\bibnamefont{Pastor-Satorras}}
  \bibnamefont{and}
  \bibinfo{author}{\bibfnamefont{A.}~\bibnamefont{Vespignani}},
  \bibinfo{journal}{Phys. Rev. Lett.} \textbf{\bibinfo{volume}{86}},
  \bibinfo{pages}{3200} (\bibinfo{year}{2001}).

\bibitem[{\citenamefont{Parshani et~al.}(2010)\citenamefont{Parshani, Carmi,
  and Havlin}}]{parshani10}
\bibinfo{author}{\bibfnamefont{R.}~\bibnamefont{Parshani}},
  \bibinfo{author}{\bibfnamefont{S.}~\bibnamefont{Carmi}}, \bibnamefont{and}
  \bibinfo{author}{\bibfnamefont{S.}~\bibnamefont{Havlin}},
  \bibinfo{journal}{Phys. Rev. Lett.} \textbf{\bibinfo{volume}{104}},
  \bibinfo{pages}{258701} (\bibinfo{year}{2010}).

\bibitem[{\citenamefont{Watts et~al.}(2005)\citenamefont{Watts, Muhamad,
  Medina, and Dodds}}]{watts05}
\bibinfo{author}{\bibfnamefont{D.~J.} \bibnamefont{Watts}},
  \bibinfo{author}{\bibfnamefont{R.}~\bibnamefont{Muhamad}},
  \bibinfo{author}{\bibfnamefont{D.~C.} \bibnamefont{Medina}},
  \bibnamefont{and} \bibinfo{author}{\bibfnamefont{P.~S.} \bibnamefont{Dodds}},
  \bibinfo{journal}{PNAS} \textbf{\bibinfo{volume}{102}},
  \bibinfo{pages}{11157} (\bibinfo{year}{2005}).

\bibitem[{\citenamefont{Kiss and Green}(2008)}]{kiss08}
\bibinfo{author}{\bibfnamefont{I.~Z.} \bibnamefont{Kiss}} \bibnamefont{and}
  \bibinfo{author}{\bibfnamefont{D.~M.} \bibnamefont{Green}},
  \bibinfo{journal}{Phys. Rev. E} \textbf{\bibinfo{volume}{78}},
  \bibinfo{pages}{048101} (\bibinfo{year}{2008}).

\bibitem[{\citenamefont{Gleeson et~al.}(2010)\citenamefont{Gleeson, Melnik, and
  Hackett}}]{gleeson10}
\bibinfo{author}{\bibfnamefont{J.~P.} \bibnamefont{Gleeson}},
  \bibinfo{author}{\bibfnamefont{S.}~\bibnamefont{Melnik}}, \bibnamefont{and}
  \bibinfo{author}{\bibfnamefont{A.}~\bibnamefont{Hackett}},
  \bibinfo{journal}{Phys. Rev. E} \textbf{\bibinfo{volume}{81}},
  \bibinfo{pages}{066114} (\bibinfo{year}{2010}).

\bibitem[{\citenamefont{Ball et~al.}(2010)\citenamefont{Ball, Sirl, and
  Trapman}}]{ball10}
\bibinfo{author}{\bibfnamefont{F.}~\bibnamefont{Ball}},
  \bibinfo{author}{\bibfnamefont{D.}~\bibnamefont{Sirl}}, \bibnamefont{and}
  \bibinfo{author}{\bibfnamefont{P.}~\bibnamefont{Trapman}},
  \bibinfo{journal}{Math. Biosci.} \textbf{\bibinfo{volume}{224(2)}},
  \bibinfo{pages}{53} (\bibinfo{year}{2010}).

\bibitem[{\citenamefont{Palla et~al.}(2009)\citenamefont{Palla, Pollner,
  Barab\'{a}si, and Vicsek}}]{gross09}
\bibinfo{author}{\bibfnamefont{G.}~\bibnamefont{Palla}},
  \bibinfo{author}{\bibfnamefont{P.}~\bibnamefont{Pollner}},
  \bibinfo{author}{\bibfnamefont{A.-L.} \bibnamefont{Barab\'{a}si}},
  \bibnamefont{and} \bibinfo{author}{\bibfnamefont{T.}~\bibnamefont{Vicsek}},
  \emph{\bibinfo{title}{Adaptive Networks}} (\bibinfo{publisher}{Springer},
  \bibinfo{year}{2009}), chap.~\bibinfo{chapter}{2}, pp.
  \bibinfo{pages}{11--50}.

\bibitem[{\citenamefont{Marceau et~al.}(2010)\citenamefont{Marceau, No\"el,
  H\'ebert-Dufresne, Allard, and Dub\'e}}]{marceau10}
\bibinfo{author}{\bibfnamefont{V.}~\bibnamefont{Marceau}},
  \bibinfo{author}{\bibfnamefont{P.-A.} \bibnamefont{No\"el}},
  \bibinfo{author}{\bibfnamefont{L.}~\bibnamefont{H\'ebert-Dufresne}},
  \bibinfo{author}{\bibfnamefont{A.}~\bibnamefont{Allard}}, \bibnamefont{and}
  \bibinfo{author}{\bibfnamefont{L.~J.} \bibnamefont{Dub\'e}},
  \bibinfo{journal}{Phys. Rev. E} \textbf{\bibinfo{volume}{82}},
  \bibinfo{pages}{036116} (\bibinfo{year}{2010}).

\end{thebibliography}

\end{document}